# A Review of Techniques to Mitigate Sybil Attacks

**Nitish Balachandran**

*Department of Computer Science and Information Systems, BITS Pilani*
*nitish.balachandran@gmail.com*

**Sugata Sanyal**

*School of Technology & Computer Science, Tata Institute of Fundamental Research, Mumbai, India*
*sanyals@gmail.com*

-----------------------------------------------------------ABSTRACT-----------------------------------------------------------------
**Any decentralised distributed network is particularly vulnerable to the Sybil attack wherein a malicious node masquerades as several different nodes, called Sybil nodes, simultaneously in an attempt to disrupt the proper functioning of the network. Such attacks may cause damage on a fairly large scale especially since they are difficult to detect and there has been no universally accepted scheme to counter them as yet. In this paper, we discuss the different kinds of Sybil attacks including those occurring in peer-to-peer reputation systems, self-organising networks and even social network systems. In addition, various methods that have been suggested over time to decrease or eliminate their risk completely are also analysed along with their modus operandi.**

## 1. INTRODUCTION

**A** Sybil attack [1] is one in which a malicious node on a network illegitimately claims to be several different nodes simultaneously. If an entity on a network does not have physical knowledge of the other entities, it will perceive them purely as informational abstractions called identities.

Sybil attacks occur when the one-to-one correspondence between an entity and its identity is violated.

They affect a number of environments and application domains in a variety of ways. For instance, the reputation system of a P2P network may be compromised as the attacker is able to favourably alter reputation scores by the use of the newly created rogue identities. In the worst case scenario, an attacker can create an infinite number of forged identities with just one physical device [11].

## 2. SPECIFIC TYPES OF SYBIL ATTACKS

There are numerous malicious applications of Sybil attacks in different environments such as those including, but not limited to, the variations enlisted below.

### 2.1 Routing

Sybil attacks can disrupt routing protocols in ad hoc networks, especially the multicast routing mechanism. Separate paths that initially seem disjoint may pass through the Sybil nodes of a single attacker. Another vulnerable concept is Geographical routing where malicious nodes may appear at more than one place at a time [4].

An attack in an ad hoc network and thus the availability of fake identities may further lead to a large scale attack such as distributed DoS, in addition to the inherently insecure routing protocols in such networks [12].

## 2.2 Tampering with Voting and Reputation Systems

In case of any environment where there is a voting scheme in place for purposes such as reporting and identifying node misbehaviour in the system, updating reputation scores and so on, a Sybil attack may be particularly dangerous. As an example, an attacker may create enough malicious identities to repeatedly report and subsequently remove legitimate nodes from the network. Alternatively, these malicious nodes can protect themselves from ever being removed as they are in collusion.

## 2.3 Fair Resource Allocation

Sybil attacks may also be used to enable the attacker to obtain an unfair and disproportionately large share of resources that were intended to be distributed amongst all nodes on the network equally. This attack denies legitimate nodes their deserved share of resources and also provides the malicious node with more avenues for other attacks.

## 2.4 Distributed Storage

File storage systems in peer-to-peer and wireless sensor networks can be compromised by the Sybil attack. This is achieved by defeating the fragmentation and replication processes in the file system. A system can be tricked into storing data into the multiple Sybil identities of the same node on the network.

## 2.5 Data Aggregation

Sensor network readings are computed by query protocols [13] in a network rather than returning the reading of each individual sensor. This is done to conserve energy. Sybil identities may be able to report incorrect sensor readings thereby influencing the overall computed aggregate. A malicious user may be able to significantly alter the aggregate with enough identities.

## 3. METHODS PROPOSED TO COUNTER SYBIL ATTACKS

Though there is no general, universally-accepted solution to the Sybil attack, a number of approaches for various combinations of environments and attacks have been proposed. Some methods mitigate the threat level of these attacks in a system to a satisfactory minimum without incurring an appreciable performance overhead. We must note that although they will not completely eliminate the possibility of the attack occurring, they are more than worthy of consideration.

Notable techniques to counter Sybil attacks are as under.

## 3.1 Trusted Certification

Certification is by far the most frequently cited solution to defeating Sybil attacks [5]. It involves the presence of a trusted certifying authority (CA) that validates the one is to one correspondence between an entity on the network and its associated identity. This centralised CA thus eliminates the problem of establishing a trust relationship between two communicating nodes. Douceur has proven that such kind of certification is the only method that may potentially eliminate Sybil attacks completely [1]. Although this approach intuitively seems like the ideal method to tackle these attacks, there are a number of implementation issues specifically about how the CA shall establish the entity-identity mapping. In real-world applications this may incur an appreciable performance cost particularly if performed manually on large scale systems.

## 3.2 Resource Testing

Resource Testing is the most commonly implemented solution to averting Sybil attacks. The basic principle is that the quantum of computing resources of each entity on the network is limited. A verifier then checks whether each identity has as many resources as the single physical device it is associated with. Any discrepancy indicates the possibility of a compromised node. Storage, computation and communication were initially proposed as resources. However, for a system such as a wireless sensor network, an attacker might have storage and computation resources in large capacities compared to resource-starved sensor nodes. Alternatively, verification messages for verifying communication resources might flood the entire system itself. Hence, all three are inadequate choices for sensor networks.

Radio resource testing, proposed by Newsome et al. in [6], is an extension of the resource testing verification method for wireless sensor networks. The key assumptions of this approach are that any physical device has only one radio and that this radio is incapable of transmitting and receiving messages on more than one channel at any given time.

Resource tests have been suggested by many as a minimal defence against Sybil attacks where the goal is to reduce their risk substantially rather than to eliminate it altogether.

### 3.3 Recurring Costs

This method is a variation of resource testing where resource tests are conducted after specific time intervals to impose a certain "cost" on the attacker that is incurred for every identity that he controls or introduces into the network.

However a number of researchers that have endorsed this method [7, 8, 9] have used computational power in their resource tests. This in itself may be inadequate in controlling the attack since a malicious user incurs only a one-time cost (for computing resources) that may be recovered via the execution of the attack itself, as pointed out by Levine et al. in [5]. In [10] the authors make use of an economic model to propose a critical value that exists for a particular combination of application domain and attacker objective. An attack is deemed successful only if ratio of the attacker's objective value to the cost per identity exceeds this critical value. They conclude that using recurring costs or fees per identity is more effective as a deterrent to Sybil attacks than a one-time resource test.

### 3.4 Privilege Attenuation

In [16], Fong considers a different kind of Sybil attack altogether – one that is distinct from others that plague peer-to-peer and reputation systems. This attack aims to create pseudonymous or fake identities in a Social Network System (SNS) and get them to collude to favourably alter the existing trust relationships in the network. These relationships are represented via a graph-theoretic relationship model that exists between the owner of a resource and a prospective accessor of the same resource and is called a social graph. Such models are common in quite a few popular Social Network Systems such as Facebook. Access control policies are as defined by the respective SNSes themselves. This concept of relationship-based access control (ReBARC) [18, 19, 20, 21] is the basis for authorization decisions in the system.

When the counterfeit identities or fake accounts in the SNS collude, they may gain the ability to access personal, sensitive and restricted user information or perform large-scale crawls on the social graph [17]. Weakly configured access control policies on SNSes render them vulnerable to such attacks [1].

To counter this threat, Fong has proposed a particular version of Denning's Principle of Privilege Attenuation or POPA that is both a necessary and sufficient condition to thwart such attacks, along with a static policy analysis for verifying POPA compliance [16].

### 3.5 Incentive-based Detection

Margolin and Levine propose a protocol in [22] called Informant that is based on an economic incentive policy and is a general solution that is not specific to any particular application domain. An entity (called the detective) rewards Sybils for revealing themselves. An identity gives the name of the target peer and a security deposit to the detective while the target peer receives the deposit and a certain reward. A Dutch auction is used to establish the minimum reward that will reveal a Sybil node. No physical tokens are required such as radios and clock skews unlike other Sybil detection approaches [3, 22, 23].

### 3.6 Location / Position Verification

This solution is specific to Wireless ad hoc Networks. Methods employing this technique make use of the fact that any identities that are projected by any single physical device must be in the same location. Locations are verified using specific methods such as triangulation [25]. So for an attacker with a single physical device, all Sybil identities will be in the same place or will appear to move together.

Tangpong et al. have proposed a solution in [24] based on the above strategy.

### 3.7 RSSI-based scheme

In [14], Demirbas and Song introduce a method for Sybil detection based on the Received Signal Strength Indicator (RSSI) of messages. The cooperation of one additional node (and hence one message communication) is required for the proper functioning of this protocol. A localisation algorithm is used in this scheme Sybil attacks can be detected with a completeness of 100% with few false positive alerts. Despite the fact that RSSI is unreliable and that transmissions via radio are non-isotropic, the use of ratios of RSSIs from multiple receivers solves this problem.

### 3.8 Random Key Predistribution

This technique enables nodes on a wireless sensor network to establish secure links for communicating with each other [15]. In random key predistribution, a set of keys are assigned at random to a node enabling it to discover or compute the common keys that it shares with its neighbouring nodes. Node-to-node secrecy is ensured by using the common keys as a shared secret session key. The main ideas are the association of the identity with the key assigned to a node and the validation of the key. Validation involves ensuring that the network is able to validate the keys that an identity might have. The forged Sybil identity will not pass the key validation test as the keys associated with a random identity will most likely, not have an appreciable intersection with the compromised key set.

Table I: Various approaches to tackle Sybil attacks in different application domains and their limitations

| S. No. | Technique to mitigate Sybil attack | Disadvantages / Limitations | Application Domain |
|--------|-----------------------------------|-----------------------------|--------------------|
| 1 | Trusted Certification | Significant performance overhead and expense [1][6][4] | General |
| 2 | Resource Testing | Ineffective for most systems [1][6][4] | General |
| 3 | Recurring Fees | Requires the use of electronic cash or of significant human effort [7] | General |
| 4 | Privilege Attenuation | Only applies to monotonic policies. Significant run-time and storage overhead for generalised extensions of the idea [16] | Social Network Systems |
| 5 | Economic Incentives | May encourage Sybil attackers that have no interest in subverting the application protocols, but that are interested in being paid to reveal their presence [26] | General |
| 6 | Location/Position Verification | Limited only to ad hoc networks | Wireless ad hoc networks |
| 7 | Received Signal Strength Indicator (RSSI) – based scheme | Does not deal with existing Sybil nodes in the network, Location calculations are costly, Limited to Sensor Networks | Sensor Networks |
| 8 | Random Key Predistribution | Limited to Sensor Networks [2] | Sensor Networks |

## 4. CONCLUSION

In this paper, we have discussed the important kinds of Sybil attacks that can be launched on various application domains. We have also listed notable methods that have been proposed over time to tackle these attacks. Further, we have elaborated on their modus operandi, advantages, and limitations. TABLE I at the end of section 3 summarizes this.

**Author's Biography**

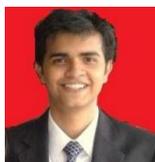

Nitish Balachandran is a final year student of Information Systems at the Birla Institute of Technology & Science (BITS), Pilani, India. His research interests include security in self-organising networks, wireless networks, the TCP/IP suite and privacy protection. He intends to pursue further studies in Networking and Security.

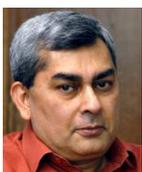

Sugata Sanyal is a Professor in the School of Technology & Computer Science at the Tata Institute of Fundamental Research, Mumbai, India (http://www.tifr.res.in/~sanyal).    He has worked in diverse areas of Computer Architecture, Parallel Processing, Fault Tolerance and Coding Theory and in the area of Security. Sanyal is in the Editorial Board of many International Journals, and is collaborating with scientists from India and abroad.